\documentstyle[epsfig,a4p,12pt]{article}

\newcommand{\etal}      {{\it et~al.}}
\newcommand{\PhysLett}  {Phys.~Lett.}

\newcommand{\NPhys}     {Nucl.~Phys.}

\newcommand{\ZPhys}     {Z.~Phys.}
\begin{document}
\begin{titlepage}
\begin{center}
{\Large  EUROPEAN ORGANIZATION FOR NUCLEAR RESEARCH}
\end{center}
\bigskip\bigskip
\begin{flushright}
{\Large  CERN-EP-2000-004}\\
3rd January 2000
\end{flushright}
\bigskip
\begin{center}{\Huge\bf  Transverse and Longitudinal}
\end{center}
\begin{center}{\Huge\bf  Bose-Einstein Correlations}
\end{center}
\begin{center}{\Huge\bf  in Hadronic Z$^0$ Decays} 
\end{center}
\bigskip\medskip
\begin{center}
{\huge The OPAL Collaboration}
\end{center}
%
%
\bigskip
%
\bigskip\bigskip
\begin{center}{\Large\bf  Abstract}\end{center}
Bose-Einstein correlations in pairs of identical charged pions 
produced in a sample of 4.3 million Z$^0$ hadronic decays are
studied as a function of the three components of the momentum 
difference, transverse (``out" and ``side")
and longitudinal with respect to 
the thrust direction of the event.
A significant difference between the transverse, 
r$_{t_{side}}$, and longitudinal, r$_l$, dimensions is observed,
indicating that the emitting source of identical pions, as observed in the
Longitudinally CoMoving System, has an elongated shape.  
This is observed with a variety of selection techniques. 
Specifically, the values of the parameters obtained by fitting the extended 
Goldhaber parametrisation to the correlation function 
${\mathrm C'}~=~{\mathrm C^{DATA}}/{\mathrm C^{MC}}$
for two-jet events, selected with the Durham algorithm and 
resolution parameter y$_{cut}$~=~0.04, are 
%
%
r$_{t_{side}}$~=~(0.809~$\pm$~0.009~({\it stat})~$^{+0.019}_{-0.032}$~
({\it syst}))~fm, r$_l$~=~(0.989~$\pm$~0.011~({\it stat})~
$^{+0.030}_{-0.015}$~({\it syst}))~fm and r$_l$/r$_{t_{side}}$~=~1.222~$\pm$
~0.027~({\it stat})~$^{+0.075}_{-0.012}$~({\it syst}).
The results are discussed in the context of a recent model of
Bose-Einstein correlations based on string fragmentation.
%
\bigskip
\begin{center}
{\Large \bf Submitted to European Physical Journal C}
\end{center}
\bigskip
\end{titlepage}
\newpage
%
%
%
%
\begin{center}{\Large        The OPAL Collaboration
}\end{center}\bigskip
\begin{center}{
G.\thinspace Abbiendi$^{  2}$,
K.\thinspace Ackerstaff$^{  8}$,
P.F.\thinspace Akesson$^{  3}$,
G.\thinspace Alexander$^{ 22}$,
J.\thinspace Allison$^{ 16}$,
K.J.\thinspace Anderson$^{  9}$,
S.\thinspace Arcelli$^{ 17}$,
S.\thinspace Asai$^{ 23}$,
S.F.\thinspace Ashby$^{  1}$,
D.\thinspace Axen$^{ 27}$
% ...
}\end{center}\bigskip
\bigskip
$^{  1}$School of Physics and Astronomy, University of Birmingham,
Birmingham B15 2TT, UK
\newline
$^{  2}$Dipartimento di Fisica dell' Universit\`a di Bologna and INFN,
I-40126 Bologna, Italy
\newline
$^{  3}$Physikalisches Institut, Universit\"at Bonn,
D-53115 Bonn, Germany
\newline
% ...
$^{  8}$CERN, European Organisation for Particle Physics,
CH-1211 Geneva 23, Switzerland
\newline
$^{  9}$Enrico Fermi Institute and Department of Physics,
University of Chicago, Chicago IL 60637, USA
\newline
% ...
$^{ 16}$Department of Physics, Schuster Laboratory, The University,
Manchester M13 9PL, UK
\newline
$^{ 17}$Department of Physics, University of Maryland,
College Park, MD 20742, USA
\newline
% ...
$^{ 22}$Department of Physics and Astronomy, Tel Aviv University,
Tel Aviv 69978, Israel
\newline
$^{ 23}$International Centre for Elementary Particle Physics and
Department of Physics, University of Tokyo, Tokyo 113-0033, and
Kobe University, Kobe 657-8501, Japan
\newline
% ...
$^{ 27}$University of British Columbia, Department of Physics,
Vancouver BC V6T 1Z1, Canada
\newline
% ...
\newpage
%
%
\section{Introduction}
%
%
Bose-Einstein correlations (BECs)~\cite{comp_bec} in pairs of 
identical bosons, mainly $\pi^{\pm}$$\pi^{\pm}$, have 
been widely studied at various energies for hadronic final states
produced by different initial states: 
e$^+$e$^-$~\cite{ee_bec,ee_adl_bec},
ep~\cite{ep_bec}, {\rm p\={p}}~\cite{pp_bec},
$\pi$p, K$^{\pm}$p~\cite{pikp_bec} and heavy ion collisions~\cite{hi_bec}.
Two-particle BECs have also been studied for K$^0_S$K$^0_S$ 
pairs~\cite{kk_bec}, 
for K$^{\pm}$K$^{\pm}$~\cite{kpkm_bec}
and, at LEP2, for pions coming from W$^+$W$^-$ decays~\cite{ww_bec}.
Genuine BECs have also been observed for three charged identical 
pions~\cite{trepi}.
\newline

BECs are manifested as enhancements in the production of identical bosons
which are close to one another in phase space.
They can be analysed in terms of the correlation function
\begin{equation}
{\mathrm C}(p_{1},p_{2})~=~\frac{\rho(p_{1},p_{2})}{\rho_0(p_{1},p_{2})},
\end{equation}
where $p_1$ and $p_2$ are the four-momenta of the two bosons,
$\rho(p_{1},p_{2})$ is the measured density of the two identical bosons and
$\rho_0(p_{1},p_{2})$ is the two-particle density in the absence of
BECs.
The choice of the reference sample used to determine 
$\rho_0(p_{1},p_{2})$ 
is crucial for the measurement. 
It should have the same properties as the sample used for
$\rho(p_{1},p_{2})$ except for the presence of BECs.
In this paper we use pairs of particles with charges of 
opposite sign as the reference sample.
This sample, however, includes pairs coming from resonance decays and 
from weakly decaying particles, like the K$^{0}_{S}$.
In addition, the correlation function obtained with this reference sample 
has to be normalized and suffers, at large four-momentum differences, from
long-range correlations due to energy, momentum and charge conservation.
We therefore also use a large sample of Monte Carlo events without the
simulation of Bose-Einstein effects in order to obtain
a correlation function which is self normalized and which has a reduced 
contamination from correlated unlike-charge pairs.
This correlation function, called ${\mathrm C'}(p_{1},p_{2})$, is used 
in the paper to obtain the reference results.
\newline

The information obtained from the shape of the correlation function may be
used to infer the space-time extent of the particle emitting region. 
Most analyses have been performed assuming a spherical emitter but
several theoretical investigations have recently treated the shape of the 
correlation function in more than one dimension~\cite{theor}.
The Lund group, in particular, has developed a model for BECs based on a 
quantum mechanical interpretation of the string area fragmentation 
probability~\cite{lund_bec}.
One of the main predictions of the model, developed for two-jet events, 
is that, since momentum components longitudinal and transverse with respect 
to the string direction (i.e. event direction) are generated by different 
mechanisms, the correlation length in the longitudinal direction is 
different from that in the transverse one.
In particular, in the so called Longitudinally CoMoving System
(LCMS), the longitudinal source dimension is predicted to be larger than the
transverse dimension.
\newline

In this paper we describe an experimental study of transverse and 
longitudinal BECs performed with the high statistics sample of hadronic events 
recorded by the OPAL detector at LEP at centre-of-mass energies at and 
near the Z$^0$ resonance.
Similar two- and three-dimensional studies have been done in e$^+$e$^-$ 
annihilations at centre-of-mass energies of 34~GeV~\cite{tasso} and 
91~GeV~\cite{L3}.
We also present the results of a unidimensional analysis of BECs.
%
%
\section{Detector and Data Selection}
%
%
A detailed description of the OPAL detector can be found in
refs.~\cite{opaldete,si}.
This analysis is based mainly on the reconstruction of charged particle
trajectories and momenta in the central tracking chambers and on energy
deposits (``clusters'') in the electromagnetic calorimeters.
All tracking systems are located inside a solenoidal magnet which
provides a uniform magnetic field of 0.435~T along the beam axis
\footnote{The coordinate system is defined so that $z$ is the
coordinate parallel to the e$^+$ and e$^-$ beams, with
positive direction along the e$^-$ beam; $r$ is the
coordinate normal to the beam axis, $\phi$ is the azimuthal angle and
$\theta$ is the polar angle with respect to +$z$.}.
The magnet is surrounded by a lead-glass electromagnetic
calorimeter and a hadron calorimeter of the sampling type. Outside
the hadron calorimeter, the detector is surrounded by a system of
muon chambers. There are similar layers of detectors in the barrel
($|\rm{cos}\theta | < 0.82$) and endcap ($|\rm{cos}\theta | > 0.81$)
regions.
The central tracking detector consists of a silicon micro-vertex
detector~\cite{si}, close to the beam pipe, and three drift chamber
devices: the vertex detector, a large jet chamber, and  surrounding
$z$-chambers.
The vertex chamber is a cylindrical drift chamber covering
a range of $|\cos\theta|<0.95$ with a resolution of
50\,$\mu$m in the $r\phi$ plane and 700\,$\mu$m in the $z$ direction.
The jet chamber is a cylindrical drift chamber with an inner radius
of 25\,cm, an outer radius
of 185\,cm, and a length of about 4\,m.
Its spatial resolution is about 135\,$\mu$m in the $r\phi$ plane
from drift time information and about 6\,cm in the $z$ direction
from charge division. 
The $z$-chambers provide a more accurate $z$
measurement, with a resolution of about 300\,$\mu$m. In combination,
the three drift chambers yield a momentum resolution of
$\sigma_{p_t}/p_t \approx \sqrt{0.02^2+(0.0015\cdot p_t)^2}$
for $|\cos(\theta)| < 0.7$, where $p_t$ is the transverse momentum
in GeV/$c$.
Electromagnetic energy is measured by lead-glass calorimeters
surrounding the solenoid magnet coil. They consist of a barrel and
two endcap arrays
with a total of 11704 lead-glass blocks covering a range of
$|\cos\theta|<0.98$.
\newline

A number of selection cuts are applied to the initial data sample, 
consisting of 4.3$\cdot$10$^6$ hadronic events from Z$^0$ decays.
In order to be considered, a charged track is required to have a 
minimum of 20 hits in the jet chamber, a minimum transverse momentum of 
150 MeV/c and a maximum momentum $p$ of 65 GeV/c.
Clusters in the electromagnetic calorimeter are used in the 
jet-finding algorithm if their energies exceed
100 MeV in the barrel, or 200 MeV in the endcaps. 
Only events that are well contained in the detector are accepted, by 
requiring that $|cos \theta_{thrust}| < 0.9$, where
$\theta_{thrust}$ is the polar angle of the thrust axis, computed
using charged tracks and electromagnetic clusters that passed the above cuts.
A second set of cuts, specific to the BEC analysis, is then applied. 
Tracks are required to have a total momentum $p < 40$ GeV/c and to come
from the interaction vertex.
Electron-positron pairs from photon conversions are rejected.
Events are selected if they contain a minimum number of five 
charged tracks and if reasonably balanced in charge, i.e. if 
$|n_{ch}^{+}~-~n_{ch}^{-}|/(n_{ch}^{+}~+~n_{ch}^{-})~\leq~0.4$,
where n$_{ch}^{+}$ and n$_{ch}^{-}$ are the number of positive and 
negative charged tracks, respectively.
The analysis is performed on the inclusive sample of all events 
passing the above cuts.
In order to compare the experimental results with the predictions 
of~\cite{lund_bec}, which are relevant to two-jet events, and to
study the dependence of the results on the ``jettyness" of the event 
the same analysis is done on samples of events defined as 
``two-jet events" by the Durham jet-finding algorithm~\cite{durham},
 with various values of the resolution parameter y$_{cut}$.
%
%
\section{The Longitudinally CoMoving System}
\begin{figure}[tb]
\vspace{-19mm}
\centerline{\epsfxsize=12cm\epsffile{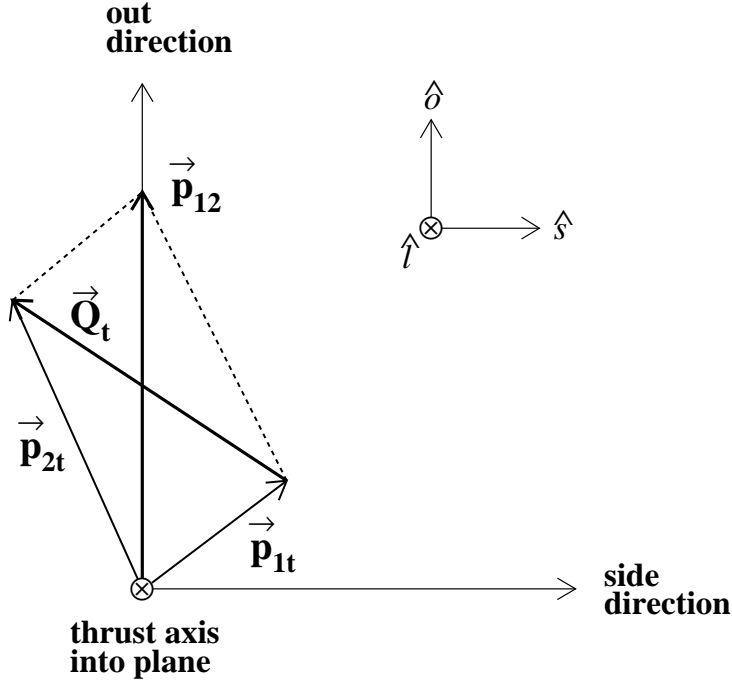}}
\vspace{-10mm}
\caption{\sl{The LCMS frame of reference drawn for a pair of particles
             where $\vec{\mathrm p}_{1t}$ and
             $\vec{\mathrm p}_{2t}$ are the projections of
             the two particles momenta onto the 
             plane perpendicular to the thrust axis.}} 
\label{lcms}
\end{figure}

%
%
In order to study transverse and longitudinal BECs, we define variables 
in the LCMS~\cite{rlcms}.
Given a pair of particles, the LCMS is the frame of reference in which 
the sum of the two particle momenta, 
$\vec{\mathrm p}_{12}$~=~($\vec{\mathrm p}_{1}$~+~
$\vec{\mathrm p}_{2}$), lies in the plane
perpendicular to the thrust axis (see Fig.~\ref{lcms}, where 
$\vec{\mathrm p}_{1t}$ and $\vec{\mathrm p}_{2t}$ 
are the projections onto the plane).
The momentum difference of the pair, $\vec{\mathrm Q} =
(\vec{\mathrm p}_{2}-\vec{\mathrm p}_{1})$, 
computed in the LCMS, is resolved into the moduli of the 
transverse component, $\vec{\mathrm Q}_{t}$, 
defined as shown in Fig.~\ref{lcms},
and of the longitudinal component
\begin{equation}
\vec{\mathrm Q}_{l} = 
|{\mathrm p}_{l_{2}}^{\prime} - {\mathrm p}_{l_{1}}^{\prime}|\hat{l}
\end{equation}
where the $\hat{l}$ direction coincides with the thrust axis.
The momentum components are marked with a prime when they are measured in 
the LCMS.
$\vec{\mathrm Q}_{t}$ may in turn be resolved into 
``out", Q$_{t_{out}}$, and ``side", Q$_{t_{side}}$, components
\begin{equation}
\vec{\mathrm Q}_{t} =
{\mathrm Q_{t_{out}}}\hat{o}~+~{\mathrm Q_{t_{side}}}\hat{s}
\end{equation}
where $\hat{o}$ and $\hat{s}$ are unit vectors in the plane perpendicular
to the thrust direction, such that 
$\vec{\mathrm p}_{12}$~=~${\mathrm p}_{12}$$\hat{o}$ defines the
``out" direction and $\hat{s}$~=~$\hat{l}$$\times$$\hat{o}$ defines the 
``side" direction.
The symbol Q is used for the invariant modulus of the four-momentum
difference {\it Q}~=~[$|{\mathrm E}_{2}-{\mathrm E}_{1}|,
(\vec{\mathrm p}_{2}-\vec{\mathrm p}_{1}$)].
\newline

It can be shown (see for instance ref.~\cite{scotto}) that, in the LCMS, 
the components Q$_{t_{side}}$ and Q$_{l}$ reflect only the difference 
in emission space of the two pions, while Q$_{t_{out}}$ depends on the 
difference in emission time as well.
Indeed, the scalar product between {\it Q} and the four-vector 
{\it P}~=~[(${\mathrm E}_{2}+{\mathrm E}_{1}),
(\vec{\mathrm p}_{2}+\vec{\mathrm p}_{1}$)] is, in the
LCMS
\begin{equation}
{\it Q}~\cdot~{\it P}~=~({\mathrm E}_{2}^{\prime}-{\mathrm E}_{1}^{\prime})
({\mathrm E}_{2}^{\prime}+{\mathrm E}_{1}^{\prime})~-~{\mathrm Q}_{t_{out}}{\mathrm p}_{12}. 
\end{equation}
Since {\it Q}~$\cdot$~{\it P}~=~0, then
\begin{equation}
{\mathrm Q}^{2}~=~({\mathrm E}_{2}^{\prime}-{\mathrm E}_{1}^{\prime})^{2}~-~
{\mathrm Q}_{{t}_{out}}^{2}
~-~{\mathrm Q}_{{t}_{side}}^{2}~-~{\mathrm Q}_{l}^{2}~=~
(
(\frac{{\mathrm p}_{12}}
{{\mathrm E}_{2}^{\prime}~+~{\mathrm E}_{1}^{\prime}})
^{2}
~-~1
)
{\mathrm Q}_{{t}_{out}}^{2}
~-~{\mathrm Q}_{{t}_{side}}^{2}~-~{\mathrm Q}_{l}^{2}.
\end{equation}
Therefore, the BECs evaluated with respect to Q$_{t_{side}}$ and Q$_{l}$ in 
the LCMS yield information on the geometrical dimensions of the pion 
emitting source.
In the string model~\cite{lund_bec} LCMS represents the local rest frame 
of the string.
In the following, we shall study the Bose-Einstein correlation function
using two different definitions of the reference sample.
%
%
\section{The Bose-Einstein Correlation function}
%
%
The three-dimensional correlation function C is defined, 
in a small phase space volume around each
triplet of Q$_{t_{out}}$, Q$_{t_{side}}$ and Q$_{l}$ values, as the number of
like-charge pairs in that volume divided by the number of 
unlike-charge pairs, used as a reference sample:
\begin{equation}
{\mathrm C}({\mathrm Q}_{t_{out}},{\mathrm Q}_{t_{side}},{\mathrm Q}_{l})~=~
\frac{N_{\pi^+ \pi^+} + N_{\pi^- \pi^-}}{N_{\pi^+ \pi^-}}~=~
\frac{N_{like}}{N_{unlike}}.
\end{equation}

Coulomb interactions between charged particles affect differently like- 
and unlike-charge pairs and thus modify the correlation function. 
A correction, based on the Gamow factors~\cite{gamow}, is applied:
each pair of like-charge pions is weighted by a factor
\begin{equation}
{\mathrm G}_l({\mathrm Q}) \ = \ (e^{2\pi \eta} - 1)/2\pi \eta \ ,
\end{equation}
where $\eta \ = \ \alpha_{\rm em} m_{\pi}/$Q; 
each pair of unlike-charge pions is weighted by a factor
\begin{equation}
{\mathrm G}_u({\mathrm Q}) \ = \ (1 - e^{- 2\pi \eta})/2\pi \eta \ .
\end{equation}

\begin{figure}[t]
\vspace{-12mm}
\centerline{\epsfxsize=10cm\epsffile{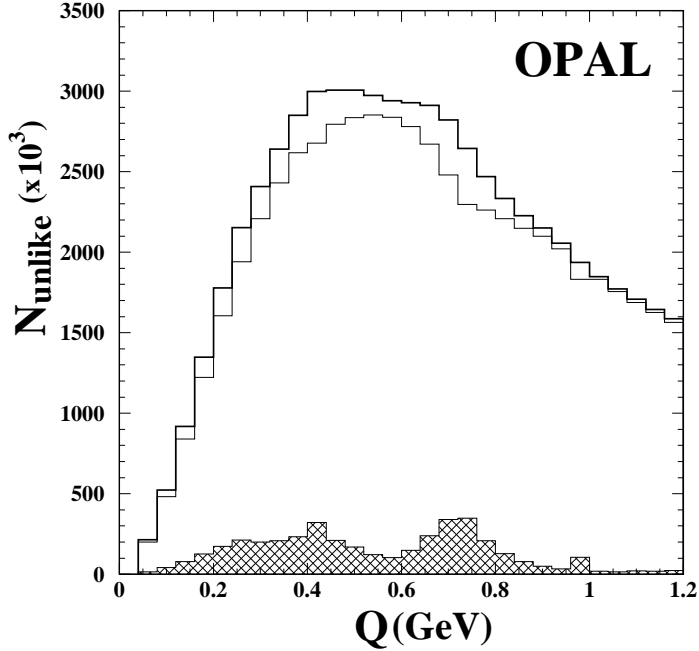}}
\caption{\sl{The unlike-charge pairs distribution vs Q before (thick line)
         and after (thin line) the subtraction of the contribution from 
         resonance decays (filled histograms; the higher peak is due to
         the $\rho^0$, the lower structure to the $\omega^0$, $\eta$ and
         K$^0_S$).}}
\label{corr}
\end{figure}
\begin{figure}[t]
\vspace{-1mm}
\mbox{\epsfxsize=8.4cm\epsffile{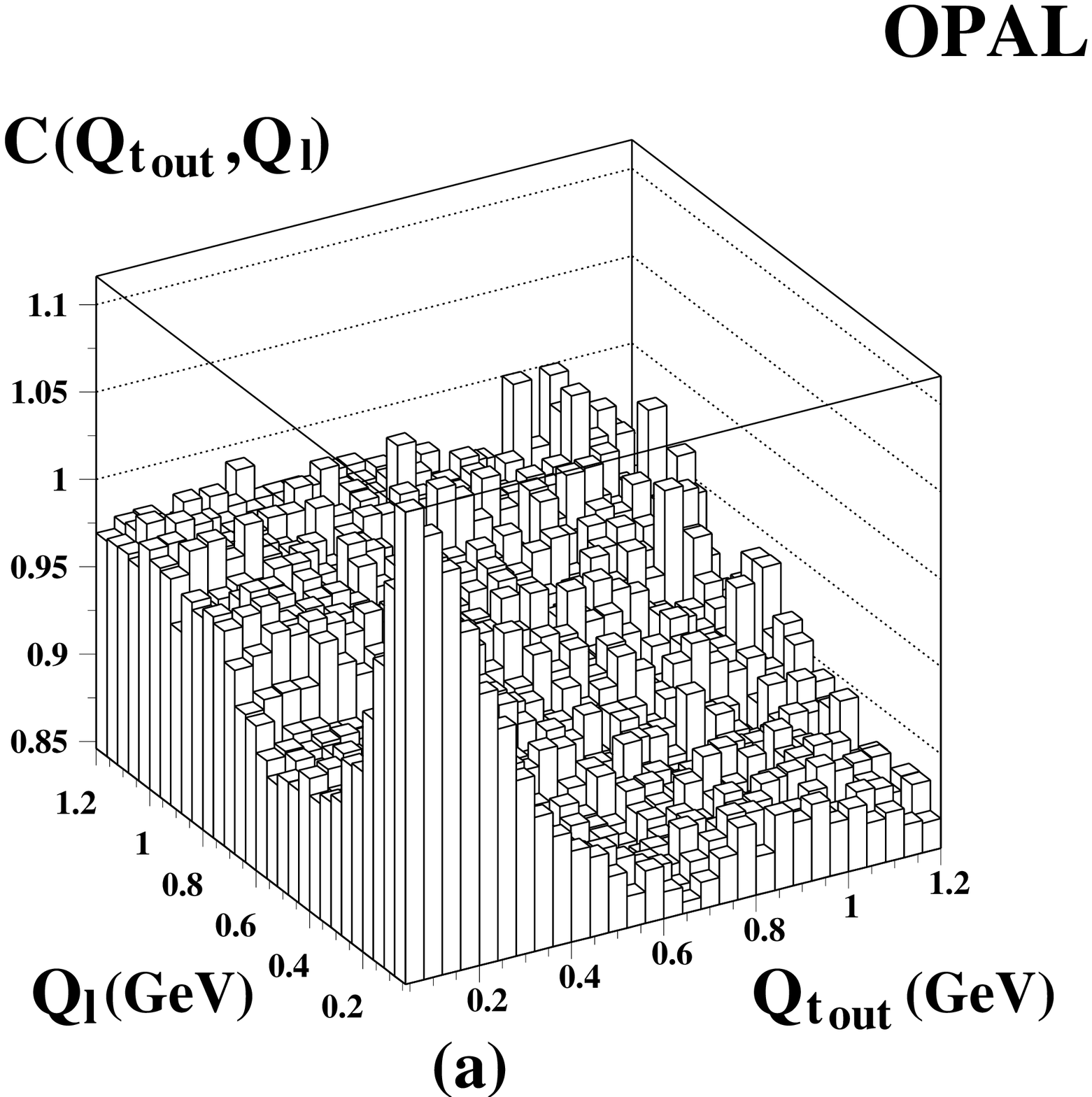}
\epsfxsize=8.4cm\epsffile{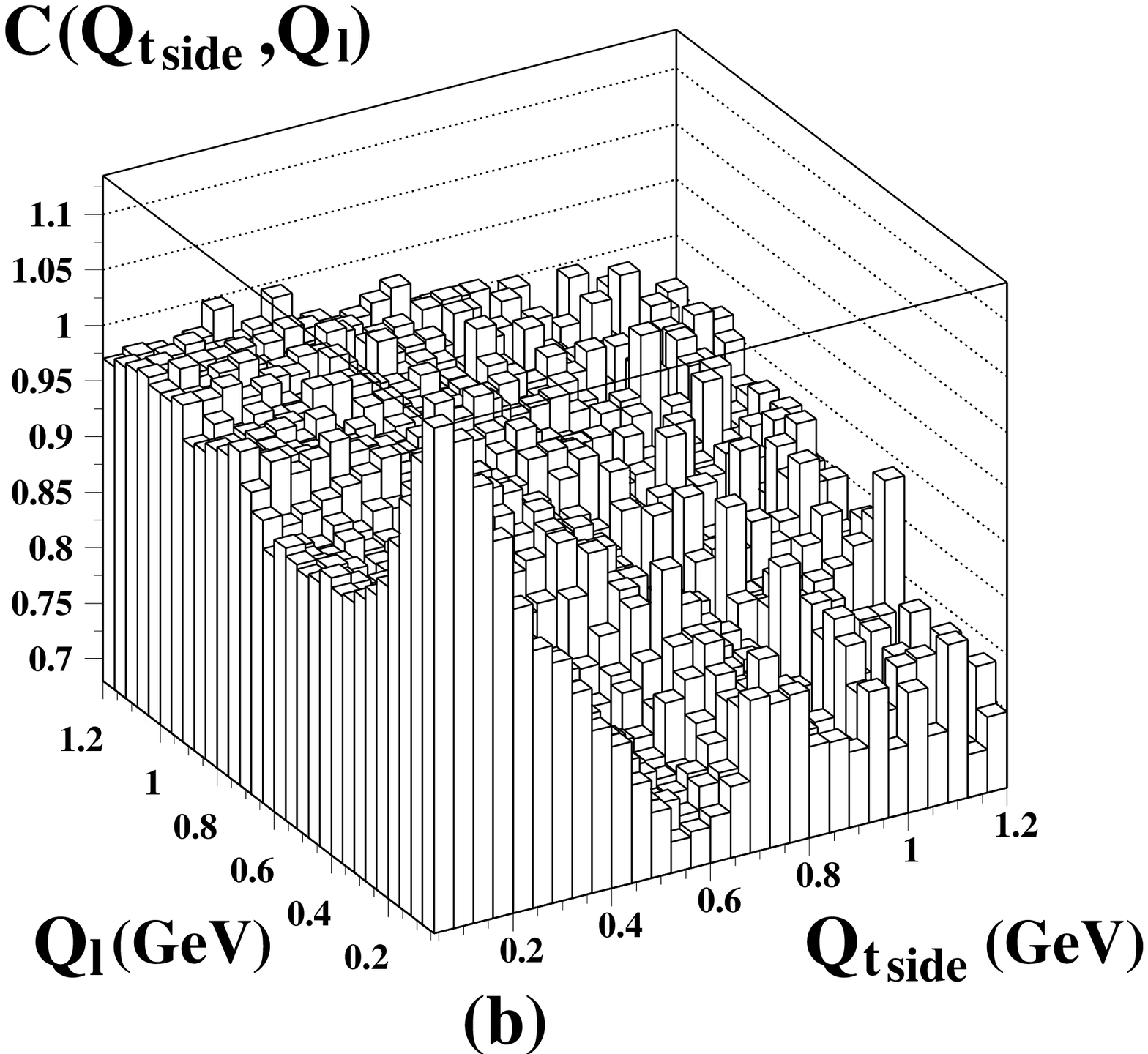}}
\caption{\sl{The projections of the three-dimensional correlation function 
             ${\mathrm C}$(Q$_{t_{out}}$,Q$_{t_{side}}$,Q$_l$),
             for two-jet events selected with y$_{cut}$~=~0.04,
             onto the (a) Q$_{t_{out}}$-Q$_l$ and the (b) Q$_{t_{side}}$-Q$_l$ 
             planes when the third component has values up to 200 MeV.
             Notice the different behaviour of ${\mathrm C}$ as function
             of Q$_{t_{side}}$ and Q$_l$.}}
\label{data}
\end{figure}

The use of unlike-charge pairs as a reference sample gives large
distortions of the correlation function in some regions of the
domain (Q$_{t_{out}}$,Q$_{t_{side}}$,Q$_{l}$), caused by the presence 
of correlated $\pi^{+}\pi^{-}$ pairs originating from decays of hadron 
resonances and of weakly decaying particles.
In order to correct the correlation function,
the Q$_{t_{out}}$, Q$_{t_{side}}$ and Q$_l$ distribution of the 
decay products of $\omega$, $\eta$, $\eta^{\prime}$, K$^0_S$,
 $\rho^0$, f$_0$(980) and f$_2$(1270) 
is determined for a sample of Jetset~7.4~\cite{Jetset} multihadronic
Monte Carlo events.
Their contribution, due mainly to
$\omega$, $\eta$, K$^0_S$ and $\rho^0$, is subtracted from the unlike 
pion distribution $N_{unlike}$.
The OPAL version of Jetset used here reproduces the resonance 
structures reasonably well, although not perfectly.
In particular the shape of the $\rho^0$ (0.64 $\leq$ Q $\leq$ 0.80~GeV) 
is not well modelled.
For each resonance, differences between the simulated and 
measured~\cite{lafferty} resonance rates are taken into 
account by scaling the distribution using the ratio of the measured 
production rate and the corresponding rate in Jetset. 
The resonance distributions obtained are then summed and scaled 
by the number of selected events. 
Figure~\ref{corr} shows the distribution $N_{unlike}$ in
the one-dimensional variable Q.
The contribution of the decay products and the same distribution
after the subtraction of this contribution are also shown.
Two-dimensional projections of the function 
C(Q$_{t_{out}}$,Q$_{t_{side}}$,Q$_l$), after the correction for 
Coulomb and resonance decay effects, are shown in Fig.~\ref{data} 
for the sample of two-jet events selected, as an example, with 
y$_{cut}$~=~0.04 and with
the third component, not plotted, limited to values up to 200~MeV. 
The histogram bin size, 40~MeV, is chosen to match the momentum 
resolution of the detector.
The presence of Bose-Einstein correlations is seen as the sharp peak 
at small values of Q$_{t_{out}}$, Q$_{t_{side}}$ and Q$_l$.
%
%
\section{Parametrisation of the Correlation function}
%
%
A minimum $\chi^2$ fit to the measured three-dimensional correlation 
function is performed using the following extended 
Goldhaber~\cite{gold} parametrisation:
\begin{equation}
{\mathrm C}({\mathrm Q}_{t_{out}},{\mathrm Q}_{t_{side}},{\mathrm Q}_{l})=
{\mathrm N}(1+\lambda e^{-({\mathrm Q}_{t_{out}}^{2}r_{t_{out}}^{2}~+~
{\mathrm Q}_{t_{side}}^{2}r_{t_{side}}^{2}~+~{\mathrm Q}_{l}^
{2}r_{l}^{2})})
{\mathrm F}({\mathrm Q}_{t_{out}},{\mathrm Q}_{t_{side}},{\mathrm Q}_{l})
\end{equation}
with
\begin{equation}
{\mathrm F}({\mathrm Q}_{t_{out}},{\mathrm Q}_{t_{side}},{\mathrm Q}_{l})=
(1+\delta_{t_{out}}{\mathrm Q}_{t_{out}}+
\delta_{t_{side}}{\mathrm Q}_{t_{side}}+
\delta_{l}{\mathrm Q}_{l}+
\epsilon_{t_{out}}{\mathrm Q}_{t_{out}}^2+
\epsilon_{t_{side}}{\mathrm Q}_{t_{side}}^2+
\epsilon_{l}{\mathrm Q}_{l}^2).
\end{equation}
The chaoticity parameter $\lambda$ measures the strength of the correlation, 
r$_{t_{out}}$, r$_{t_{side}}$ and r$_l$ indicate the transverse and 
longitudinal extent of the two-pion source, N is a normalization 
factor necessary since the reference sample $N_{unlike}$ is not 
normalized to the sample of like-charge pairs.
The term 
${\mathrm F}({\mathrm Q}_{t_{out}},{\mathrm Q}_{t_{side}},{\mathrm Q}_{l})$
accounts for long-range two-particle correlations, due to energy and
charge conservation and to phase space constraints.
Alternative forms for this function have been considered, as in particular
a fit with only the linear long-range terms $\delta_{i}{\mathrm Q}_{i}$ 
({\it i}~=~$t_{out}$,~$t_{side}$,$l$), and the results did not change 
significantly. The best results (lower values of $\chi^2$/DoF and 
stability) are obtained with formula (10).
The fits are performed over the range 
0.04~$\leq~{\mathrm Q}_{t_{out}},{\mathrm Q}_{t_{side}},
{\mathrm Q}_l~\leq$~1.2 GeV.
The region below 0.04~GeV is excluded because of the limited 
momentum resolution at low Q$_i$ values and of the presence 
of residual photon conversion pairs.
Even after the subtraction procedure described in section 4, a few
regions show significant distortions (see Fig.~3)
assumed to be due to residual effects of pairs
coming from resonance decays: these regions, corresponding to 
$0.28 \leq \sqrt{ {{\mathrm Q}_{t_{out}}}^2+{{\mathrm Q}_{t_{side}}}^2
+{{\mathrm Q}_l}^2} \leq 0.44$ GeV and 
$0.64 \leq \sqrt{ {{\mathrm Q}_{t_{out}}}^2+{{\mathrm Q}_{t_{side}}}^2
+{{\mathrm Q}_l}^2} \leq 0.80$ GeV, are not used in the fits.
\begin{table}[htbp]
\begin{center}
\footnotesize
\begin{tabular}{|l||c|c|c|c|c|}
\hline
Parameter                  & y$_{cut}$~=~0.01  & y$_{cut}$~=~0.02  &
y$_{cut}$~=~0.04  & y$_{cut}$~=~0.06   & inclusive sample
\cr
\hline\hline
N                          & $0.803 \pm 0.002$ & $0.815 \pm 0.001$ &
$0.825 \pm 0.001$ & $0.831 \pm 0.001$ & $0.842 \pm 0.001$ 
\cr
$\lambda$                  & $0.479 \pm 0.005$ & $0.464 \pm 0.004$ &
$0.454 \pm 0.004$ & $0.446 \pm 0.004$ & $0.442 \pm 0.004$
\cr
$r_{t_{out}}$~[fm]         & $0.522 \pm 0.007$ & $0.520 \pm 0.007$ &
$0.525 \pm 0.006$ & $0.523 \pm 0.006$ & $0.536 \pm 0.006$
\cr
$r_{t_{side}}$~[fm]        & $0.750 \pm 0.007$ & $0.767 \pm 0.006$ &
$0.783 \pm 0.006$ & $0.787 \pm 0.006$ & $0.809 \pm 0.006$
\cr
$r_l$~[fm]                 & $1.013 \pm 0.011$ & $1.014 \pm 0.010$ &
$1.015 \pm 0.009$ & $1.011 \pm 0.009$ & $1.018 \pm 0.009$
\cr
$\delta_{t_{out}}$~[GeV$^{-1}$]  & $-0.015 \pm 0.004$ & $-0.010 \pm 0.004$ &
$-0.003 \pm 0.003$ & $-0.002 \pm 0.003$ & $0.004 \pm 0.003$
\cr
$\delta_{t_{side}}$~[GeV$^{-1}$] & $-0.149 \pm 0.004$ & $-0.148 \pm 0.004$ &
$-0.146 \pm 0.003$ & $-0.145 \pm 0.003$ & $-0.142 \pm 0.003$
\cr
$\delta_l$~[GeV$^{-1}$]          & $0.300 \pm 0.005$ & $0.262 \pm 0.004$ &
$0.228 \pm 0.004$ & $0.212 \pm 0.004$ & $0.178 \pm 0.003$
\cr
$\epsilon_{t_{out}}$~[GeV$^{-2}$]  & $0.001 \pm 0.004$ & $0.010 \pm 0.003$ &
$0.013 \pm 0.003$ & $0.015 \pm 0.003$ & $0.015 \pm 0.002$
\cr
$\epsilon_{t_{side}}$~[GeV$^{-2}$] & $-0.032 \pm 0.004$ & $0.000 \pm 0.004$ &
$0.022 \pm 0.003$ & $0.034 \pm 0.003$ & $0.053 \pm 0.003$
\cr
$\epsilon_l$~[GeV$^{-2}$]          & $-0.072 \pm 0.004$ & $-0.055 \pm 0.003$ &
$-0.039 \pm 0.003$ & $-0.031 \pm 0.003$ & $-0.016 \pm 0.002$
\cr
\hline
$\chi^2$/DoF                     & $30089/24428$ & $31529/24428$ &
$32758/24428$   & $33261/24428$   & $34632/24428$
\cr
\hline
$r_l/r_{t_{side}}$               & $1.351 \pm 0.027$ & $1.322 \pm 0.025$ &
$1.296 \pm 0.021$ & $1.285 \pm 0.021$ &  $1.258 \pm 0.020$ 
\cr
\hline
\end{tabular}
\caption{\sl{Results of the fits of Eq.~9 to the measured three-dimensional 
         correlation function 
         ${\mathrm C}$({\rm Q}$_{t_{out}}$,{\rm Q}$_{t_{side}}$,{\rm Q}$_{l}$)
         over the range 0.04~$\leq$~
         {\rm Q}$_{t_{out}}$,{\rm Q}$_{t_{side}}$,{\rm Q}$_{l}$
         ~$\leq$~1.2 GeV, excluding the regions affected by residual
         resonance decay products described in the text.
         The quoted errors are statistical uncertainties obtained 
         from the fits. The quality of the fits is indicated
         by the value of $\chi^2$/DoF.}}
\label{tab_fit}
\end{center}
\end{table}

The values of the parameters resulting from the fits to the correlation
functions for two-jet events (with different values of y$_{cut}$)
and for the inclusive sample are given in Table~\ref{tab_fit}.
For all the fits, the transverse and longitudinal radii are significantly
different.
In particular, the ratio r$_{l}$/r$_{t_{side}}$ between the longitudinal
and transverse source radii is
r$_{l}$/r$_{t_{side}}$~=~$1.296~~\pm~0.021$~({\it stat})
for two-jet events selected with y$_{cut}$~=~0.04 and
r$_{l}$/r$_{t_{side}}$~=~$1.258~\pm~0.020$~({\it stat}) for the inclusive
sample of events.
%
%
%
\section{The Correlation function C$'$~=~C$^{DATA}$/C$^{MC}$} 
%
%
In order to improve the quality of the Goldhaber fits, reducing the effects 
from long-range correlations and resonance decay products, it is usual to 
define the following ratio of correlation functions, using data and Monte 
Carlo events:
\begin{equation}
{\mathrm C}'({\mathrm Q}_{t_{out}},{\mathrm Q}_{t_{side}},{\mathrm Q}_{l})~=~
\frac{{\mathrm C}^{DATA}}{{\mathrm C}^{MC}}~=~
\frac{N^{DATA}_{like}/N^{DATA}_{unlike}}{N^{MC}_{like}/N^{MC}_{unlike}}.    
\end{equation}
\begin{figure}[t]
\mbox{\epsfxsize=8.4cm\epsffile{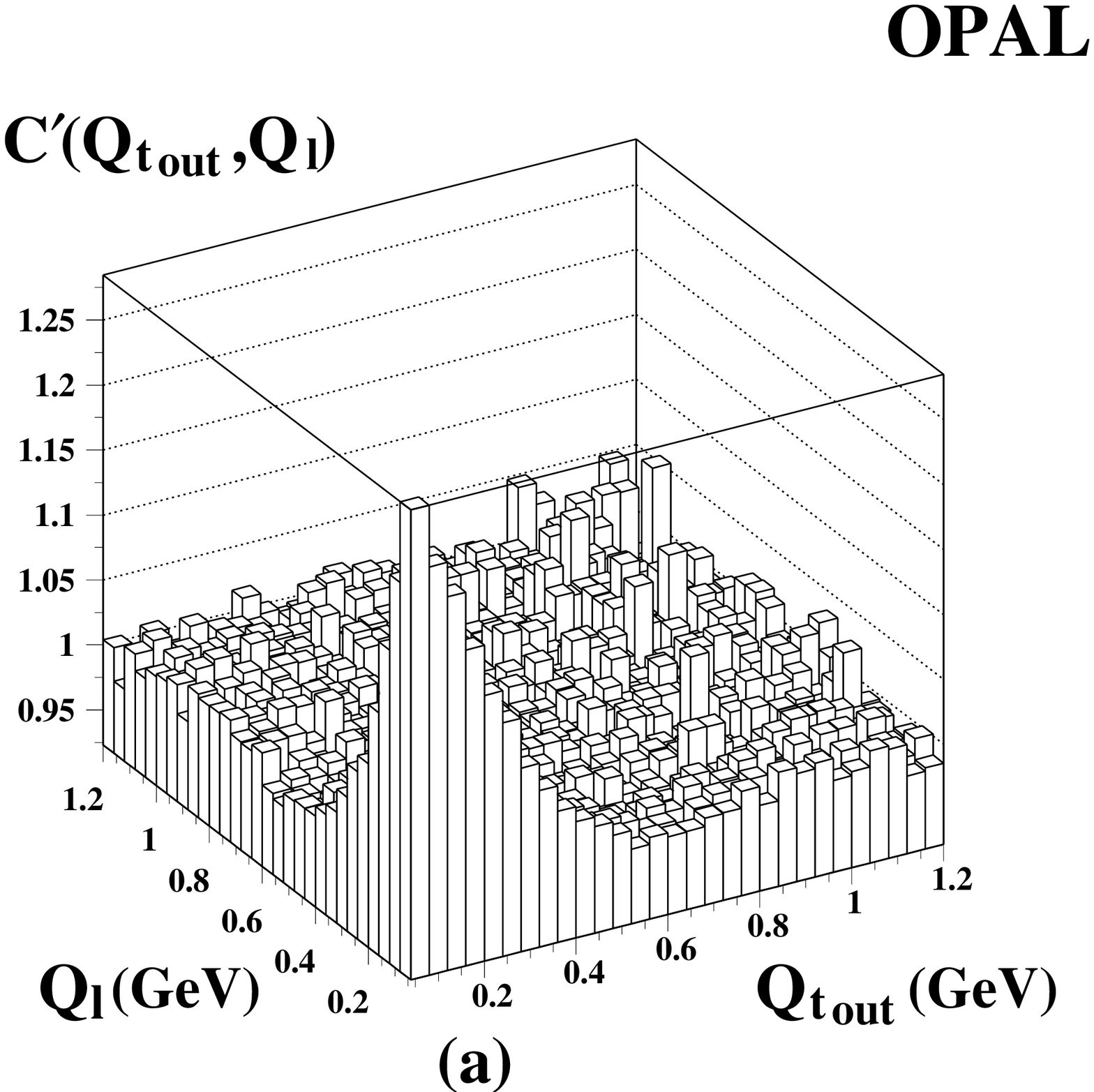}
\epsfxsize=8.4cm\epsffile{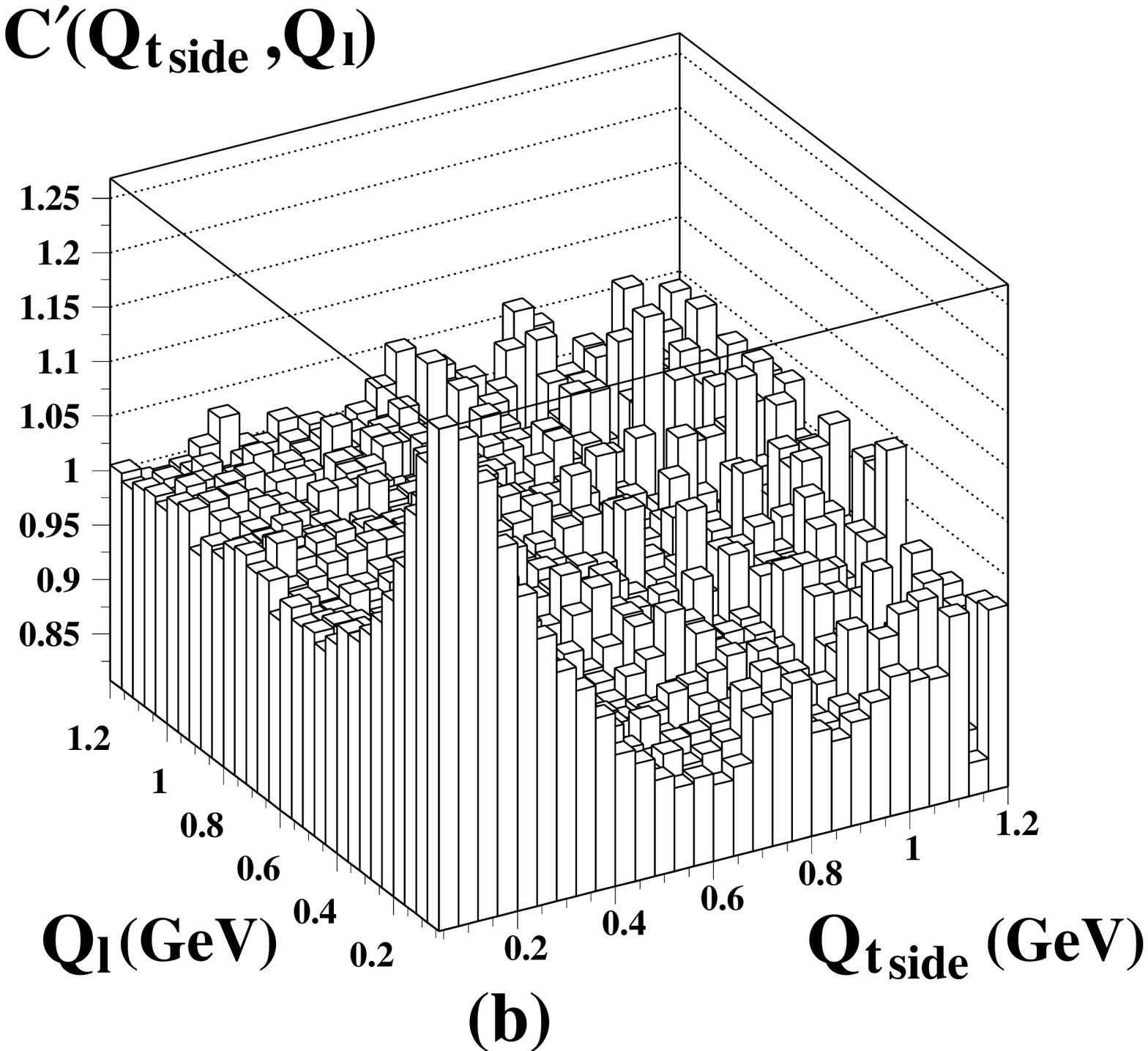}}
\caption{\sl{The projections of the three-dimensional correlation function
             ${\mathrm C'}$(Q$_{t_{out}}$,Q$_{t_{side}}$,Q$_l$),
             for two-jet events selected with y$_{cut}$~=~0.04,
             onto the (a) Q$_{t_{out}}$-Q$_l$ and the (b) Q$_{t_{side}}$-Q$_l$ 
             planes when the third component has values up to 200 MeV.
             Notice the different behaviour of ${\mathrm C'}$ as function
             of Q$_{t_{side}}$ and Q$_l$.}}
\label{data_mc}
\end{figure}
\begin{figure}[t]
\vspace{-1mm}
\mbox{\epsfxsize=8.2cm\epsffile{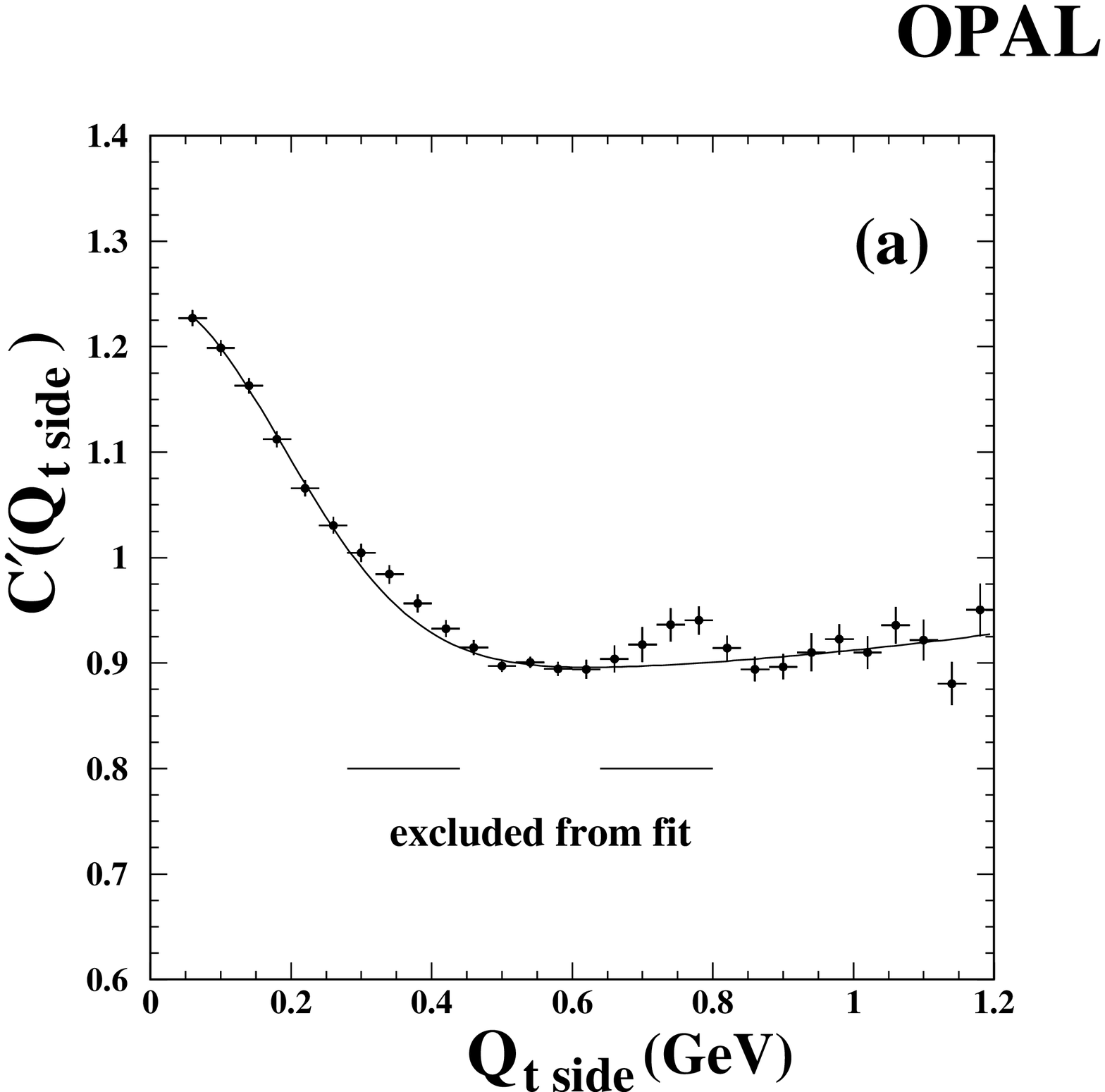}
\epsfxsize=8.2cm\epsffile{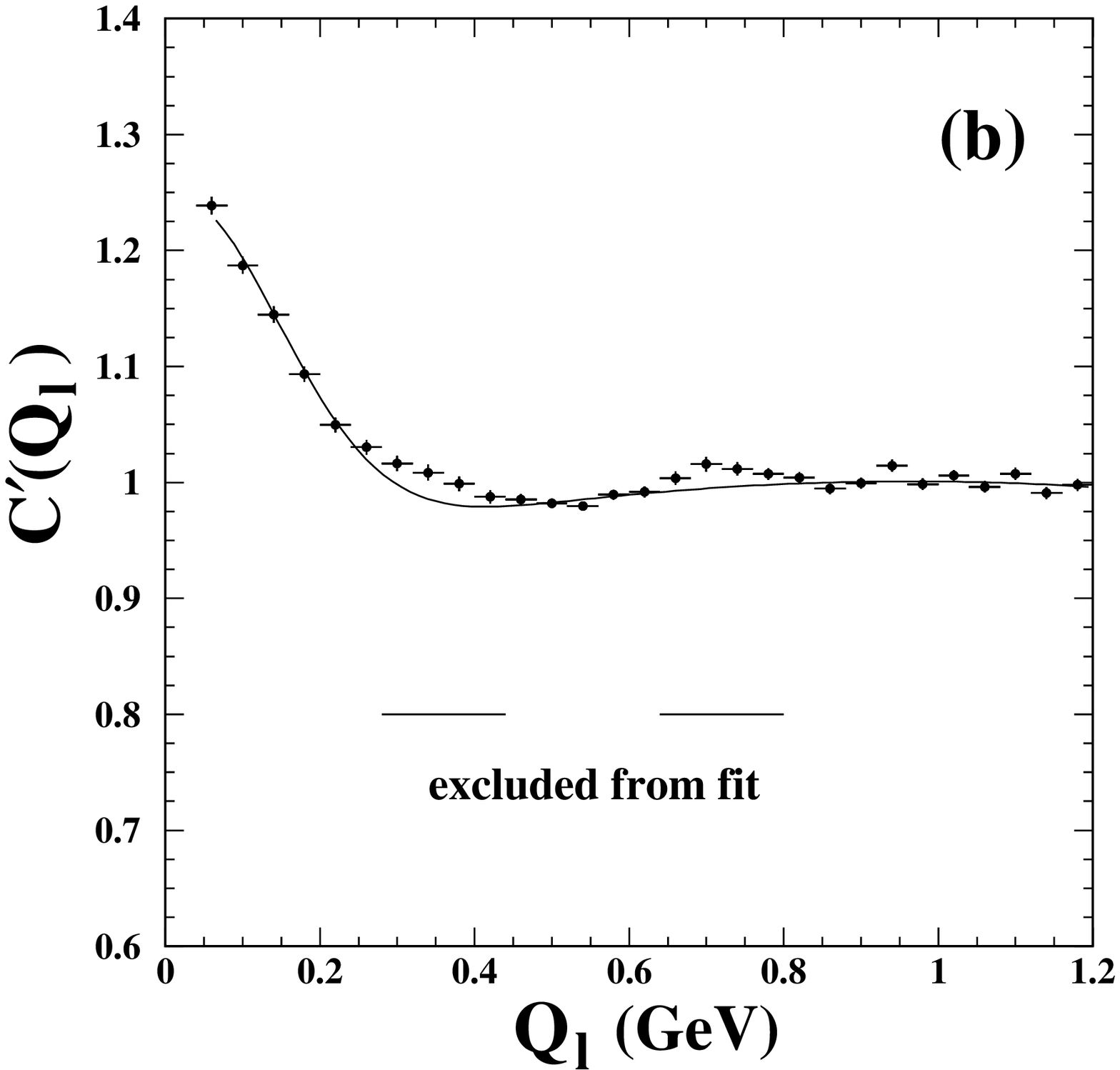}}
\caption{\sl{The projections of the three-dimensional correlation function
             ${\mathrm C'}$(Q$_{t_{out}}$,Q$_{t_{side}}$,Q$_l$),
             for two-jet events selected with y$_{cut}$~=~0.04,
             onto the (a) Q$_{t_{side}}$ and (b) Q$_l$ axes
             when the two other components have values up to 200 MeV.}}
\label{1-dim}
\end{figure}
For this purpose, a sample of 7.2 million Jetset~7.4 multihadronic
Monte Carlo events, which does not include BEC effects, is used.
The Monte Carlo simulates many of the dynamical correlations present
in the real data but not the Coulomb effect.
Therefore, in Eq.~11, only N$^{DATA}_{like}$ and N$^{DATA}_{unlike}$
are corrected by the Gamow factors given in (7) and (8).
The simulation includes the resonance decay products.
The correction, due to the differences in the measured and simulated 
resonance rates, is done by subtracting (adding) to the unlike-charge pairs
sample in the Monte Carlo the fraction of pairs simulated in excess (deficit).
Two- and one-dimensional projections of the correlation function
${\mathrm C'}$, for the sample of two-jet events selected with
y$_{cut}$~=~0.04, are shown in Fig.~\ref{data_mc} and in Fig.~\ref{1-dim},
respectively.
The Bose-Einstein correlations are clearly visible at small
${\mathrm Q}_{t_{out}}$, ${\mathrm Q}_{t_{side}}$ and Q$_l$.
The fits of Eq.~9, yield the parameters given in Table~\ref{mc}.
As can be seen in the Table, the correlation function ${\mathrm C'}$ is 
almost normalized;
 the slight difference from 1.0 is due to the difference in the average
multiplicity between data and MonteCarlo.
The $\chi^2$/DoF values for these fits are closer to unity than for those
relative to the correlation function~C.
\begin{table}[htbp]
\begin{center}
\footnotesize
\begin{tabular}{|l||c|c|c|c|c|}
\hline
Parameter                  & y$_{cut}$~=~0.01  & y$_{cut}$~=~0.02  &
y$_{cut}$~=~0.04  & y$_{cut}$~=~0.06  & inclusive sample 
\cr
\hline\hline
N                          & $0.947 \pm 0.002$ & $0.950 \pm 0.002$ &
$0.952 \pm 0.002$ & $0.954 \pm 0.001$ & $0.957 \pm 0.001$
\cr
$\lambda$                  & $0.457 \pm 0.006$ & $0.446 \pm 0.005$ &
$0.443 \pm 0.005$ & $0.441 \pm 0.005$ & $0.437 \pm 0.004$
\cr
$r_{t_{out}}$~[fm]         & $0.678 \pm 0.013$ & $0.654 \pm 0.012$ &
$0.647 \pm 0.011$ & $0.637 \pm 0.010$ & $0.621 \pm 0.010$
\cr
$r_{t_{side}}$~[fm]        & $0.781 \pm 0.010$ & $0.794 \pm 0.009$ &
$0.809 \pm 0.009$ & $0.814 \pm 0.008$ & $0.831 \pm 0.008$
\cr
$r_l$~[fm]                 & $0.987 \pm 0.013$ & $0.989 \pm 0.012$ &
$0.989 \pm 0.011$ & $0.989 \pm 0.010$ & $0.992 \pm 0.010$
\cr
$\delta_{t_{out}}$~[GeV$^{-1}$]  & $-0.004 \pm 0.005$ & $-0.002 \pm 0.004$ &
$0.006 \pm 0.004$ & $0.005 \pm 0.004$ & $0.008 \pm 0.003$
\cr
$\delta_{t_{side}}$~[GeV$^{-1}$] & $-0.062 \pm 0.005$ & $-0.071 \pm 0.004$ &
$-0.075 \pm 0.004$ & $-0.077 \pm 0.004$ & $-0.079 \pm 0.003$
\cr
$\delta_l$~[GeV$^{-1}$]          & $0.114 \pm 0.005$ & $0.103 \pm 0.005$ &
$0.092 \pm 0.004$ & $0.086 \pm 0.004$ & $0.073 \pm 0.003$
\cr
$\epsilon_{t_{out}}$~[GeV$^{-2}$]  & $0.012 \pm 0.004$ & $0.016 \pm 0.004$ &
$0.012 \pm 0.003$ & $0.013 \pm 0.003$ & $0.011 \pm 0.003$
\cr
$\epsilon_{t_{side}}$~[GeV$^{-2}$] & $-0.020 \pm 0.005$ & $0.004 \pm 0.004$ &
$0.019 \pm 0.004$ & $0.027 \pm 0.004$ & $0.037 \pm 0.003$
\cr
$\epsilon_l$~[GeV$^{-2}$]          & $-0.045 \pm 0.004$ & $-0.039 \pm 0.004$ &
$-0.032 \pm 0.003$ & $-0.028 \pm 0.003$ & $-0.021 \pm 0.003$
\cr
\hline
$\chi^2$/DoF                     & $24654/24428$ & $25249/24428$ &
$25398/24428$   & $25482/24428$   & $25836/24428$
\cr
\hline
$r_l/r_{t_{side}}$               & $1.264 \pm 0.033$ & $1.246 \pm 0.029$ & 
$1.222 \pm 0.027$ & $1.215 \pm 0.024$ &  $1.194 \pm 0.024$ 
\cr
\hline
\end{tabular}   
\caption{\sl{Results of the fits of Eq.~9 to the measured three-dimensional 
         correlation function 
         ${\mathrm C'}$({\rm Q}$_{t_{out}}$,{\rm Q}$_{t_{side}}$,{\rm Q}$_{l}$) 
         over the range 0.04~$\leq$~
         {\rm Q}$_{t_{out}}$,{\rm Q}$_{t_{side}}$,{\rm Q}$_{l}$
         ~$\leq$~1.2 GeV, excluding the regions affected by residual
         resonance decay products described in the text.
         The quoted errors are the statistical uncertainties obtained
         from the fits. The quality of the fits is indicated
         by the value of $\chi^2$/DoF.}}
\label{mc}
\end{center}
\end{table}
The dependences of $r_{t_{out}}$, $r_{t_{side}}$, $r_l$ and of the ratios
$r_l$/$r_{t_{out}}$, $r_l$/$r_{t_{side}}$ on the jet resolution parameter
y$_{cut}$ are shown in Fig.~\ref{ycut_datamc}, where the results for 
inclusive events are also presented. The main features of the results 
are a very slight decrease and increase, respectively, of $r_{t_{out}}$ 
and $r_{t_{side}}$ as y$_{cut}$ increases, while $r_l$ is independent of the 
y$_{cut}$.
The ratio $r_l$/$r_{t_{out}}$ increases slightly,
while $r_l$/$r_{t_{side}}$ decreases when y$_{cut}$ increases. 
The value of the chaoticity parameter $\lambda$ is between 0.457
and 0.437 (decreasing slowly with y$_{cut}$).
\newline

The systematic uncertainties on the parameter values are estimated 
considering the deviations with respect to a reference analysis, chosen
to be the fit of Eq.~9 performed
to the correlation function ${\mathrm C'}$(Q$_{t_{out}}$,Q$_{t_{side}}$,Q$_l$)
for two-jet events selected with y$_{cut}$~=~0.04.
In this case we have r$_l$/r$_{t_{side}}$~=~1.222~$\pm$~0.027~({\it stat}).
\begin{figure}[t]
\vspace{-25mm}
\centerline{\mbox{\epsfxsize=8cm\epsffile{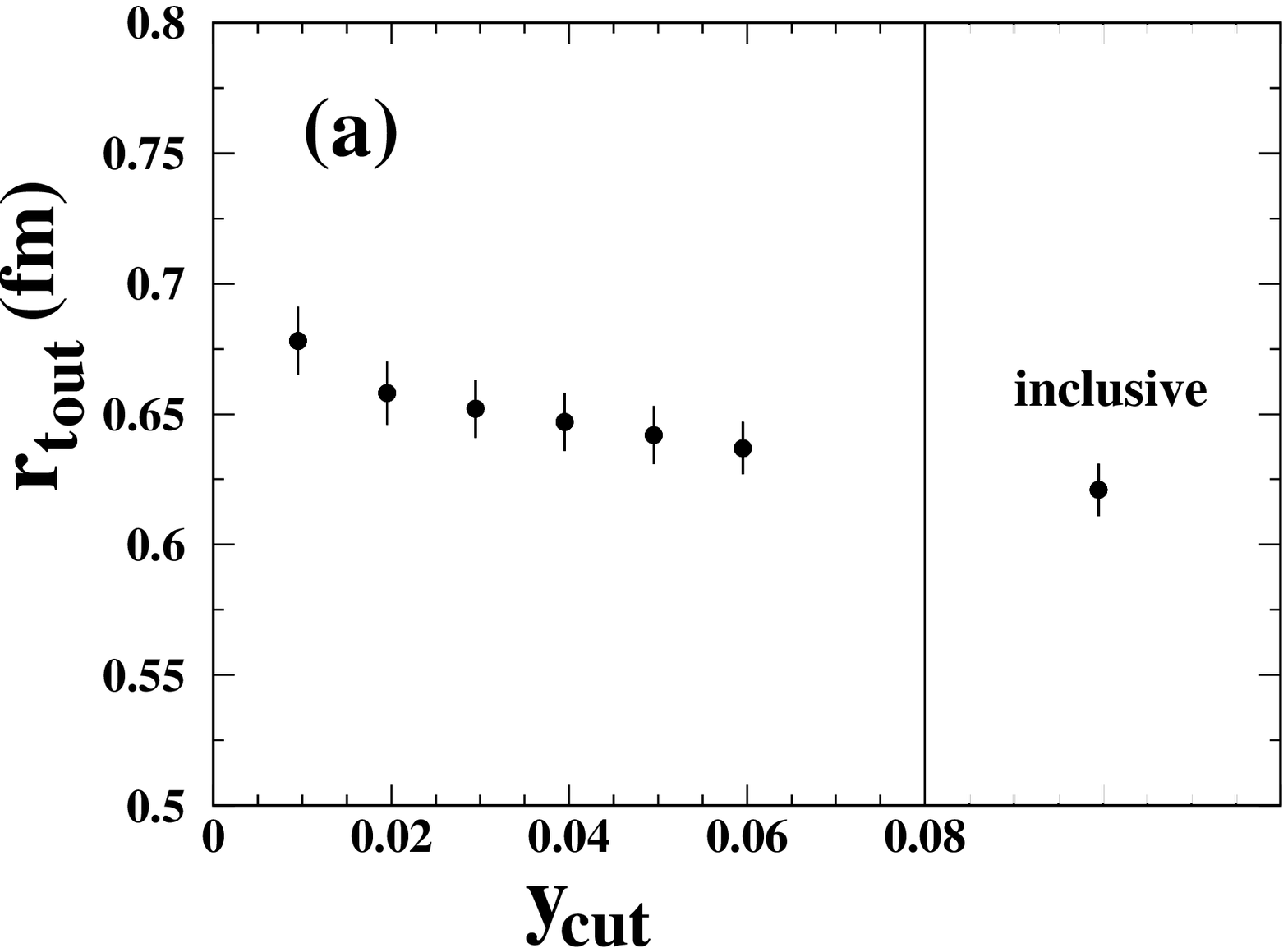}
\epsfxsize=8cm\epsffile{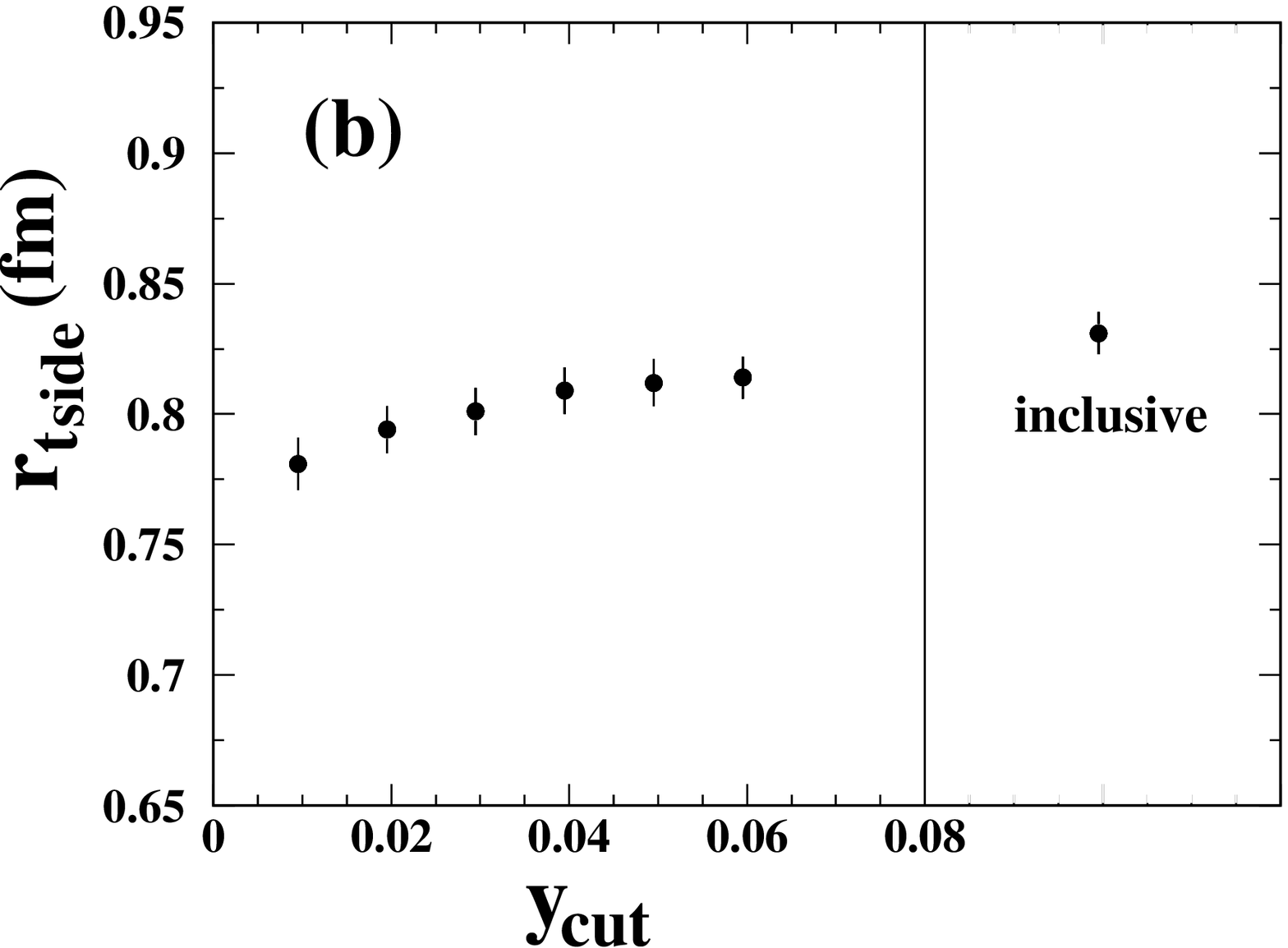}}}
\vspace{-12mm}
\centerline{\epsfxsize=8cm\epsffile{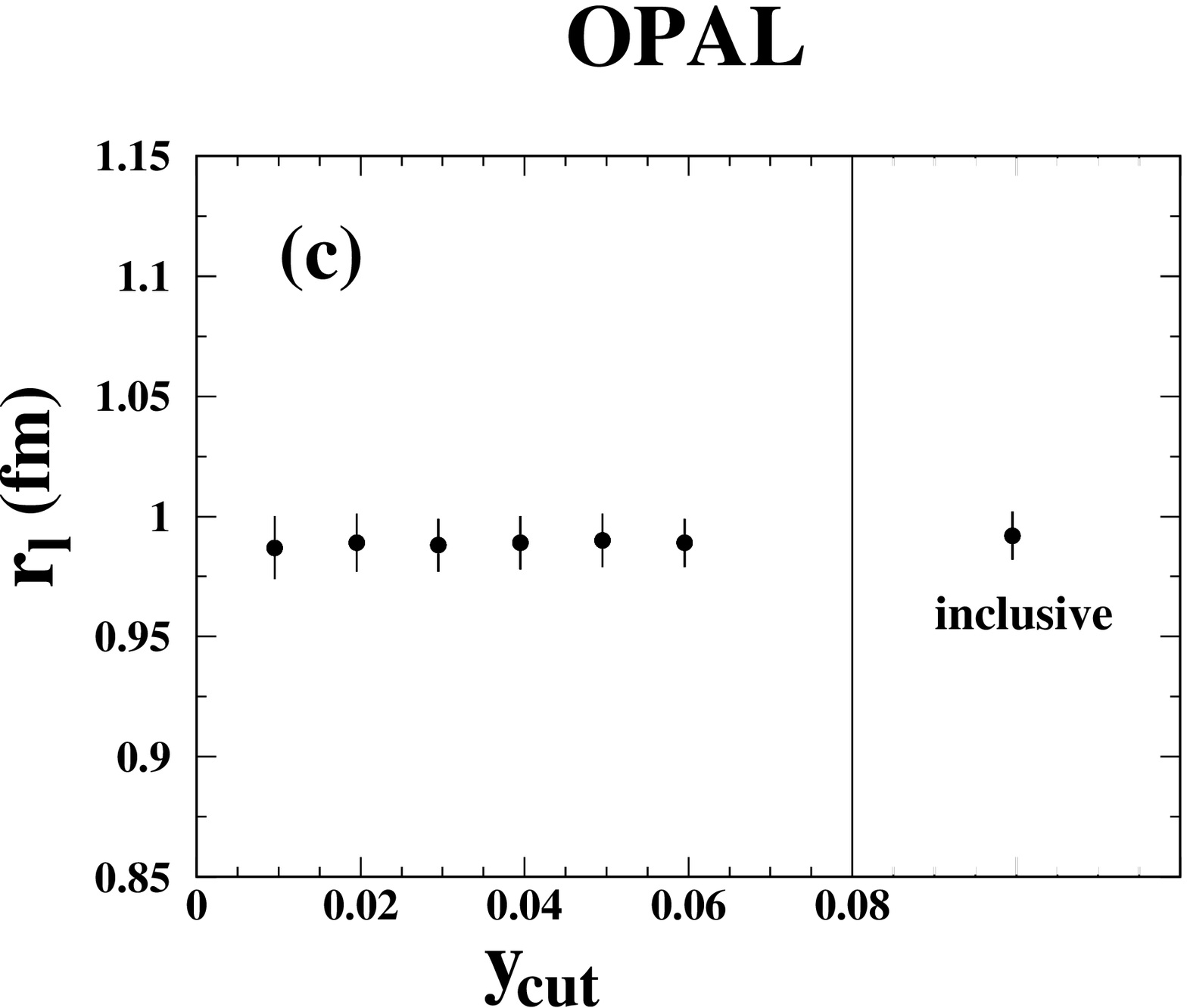}}
\vspace{-19mm}
\centerline{\mbox{\epsfxsize=8cm\epsffile{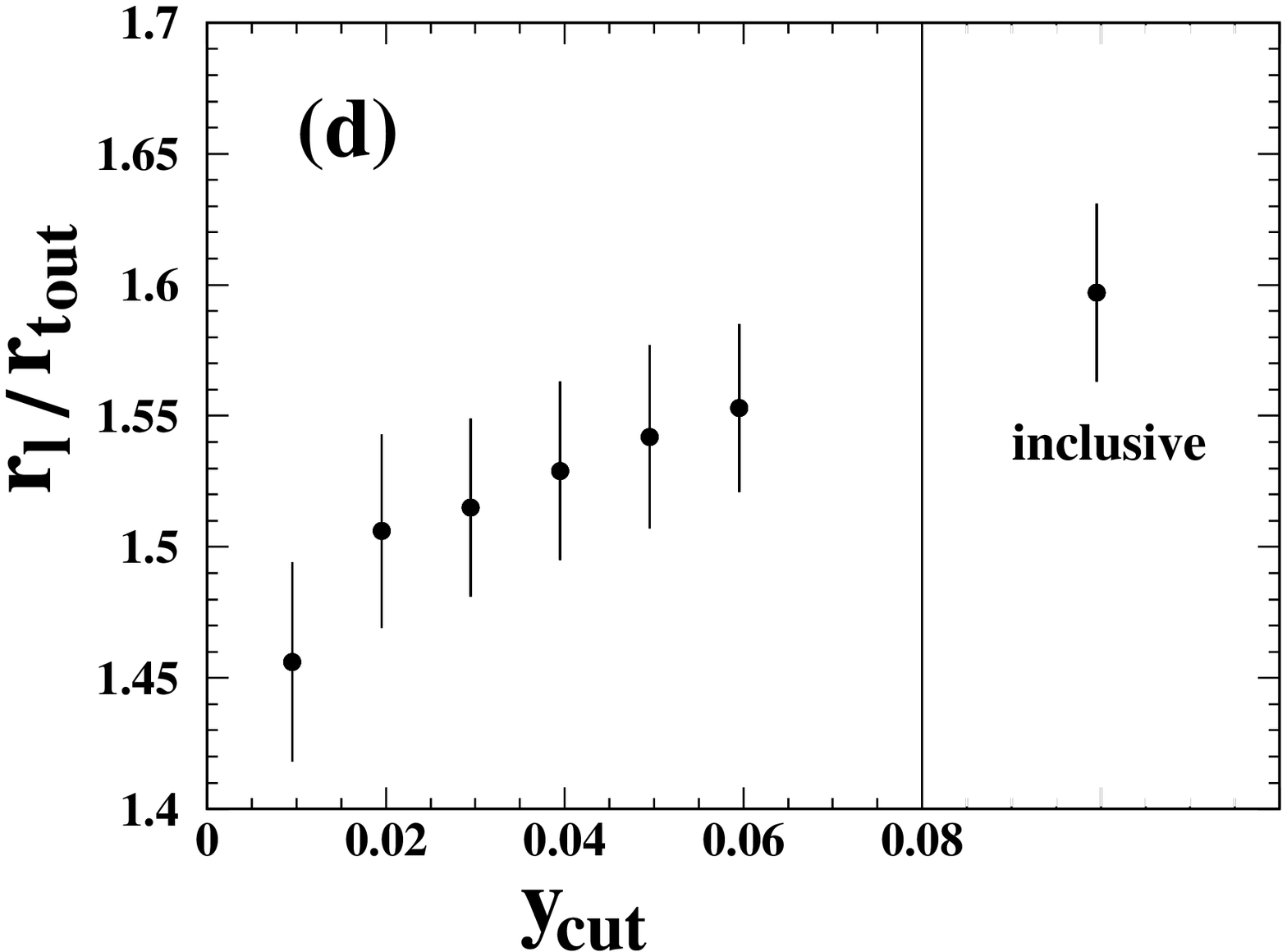}
\epsfxsize=8cm\epsffile{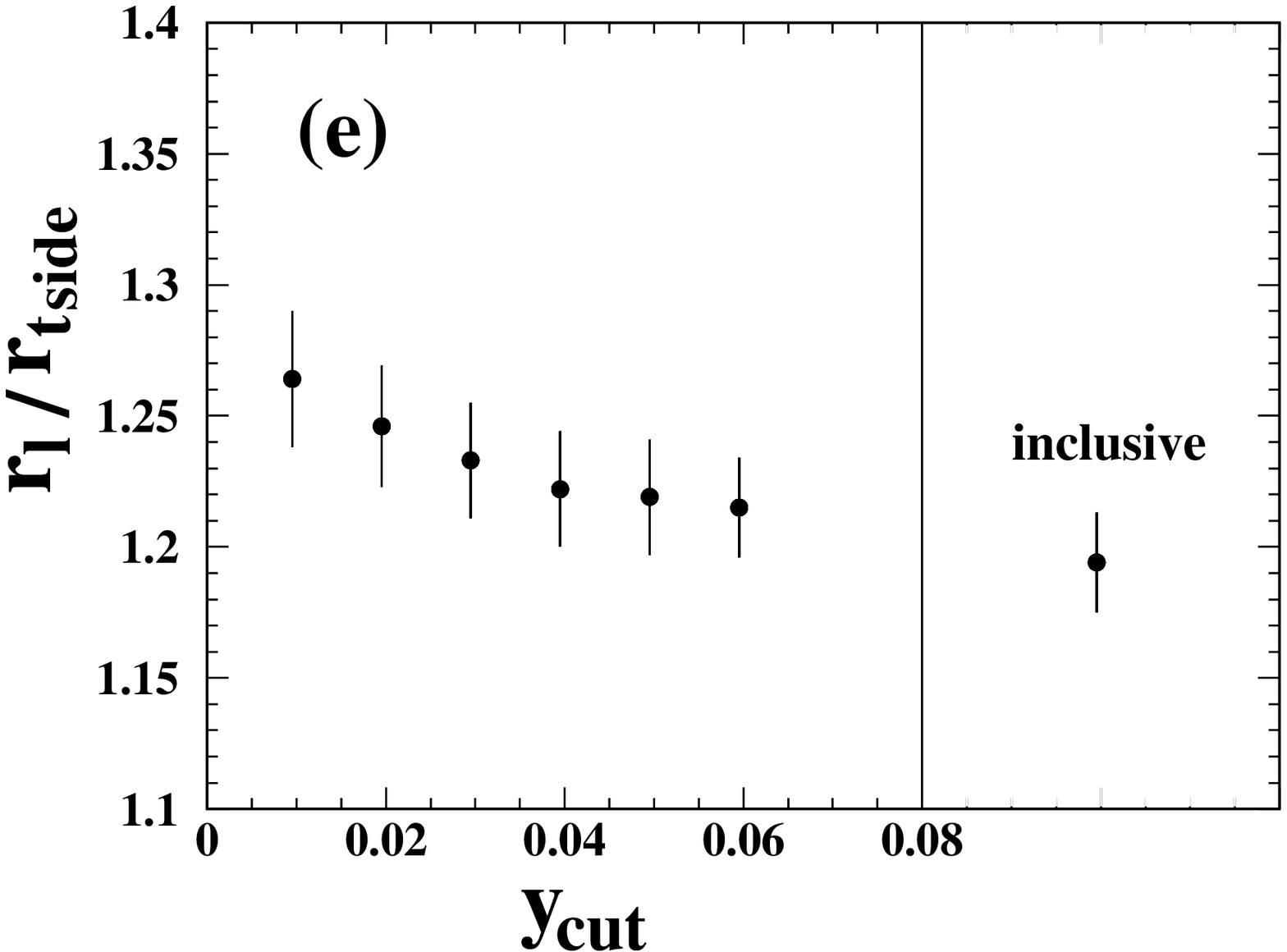}}}
\caption{\sl{Dependence on the resolution parameter y$_{cut}$
             of (a)~r$_{t_{out}}$, (b)~r$_{t_{side}}$ and (c)~r$_l$
             and of the ratios (d)~r$_l$/r$_{t_{out}}$ and
             (e)~r$_l$/r$_{t_{side}}$ for the correlation function
             ${\mathrm C'}$(Q$_{t_{out}}$,Q$_{t_{side}}$,Q$_l$).}}
\label{ycut_datamc}
\end{figure}
The analysis is repeated changing some of the selection cuts:
a maximum total momentum $p < 30$ GeV/c instead of $p < 40$ GeV/c
and a charge unbalance smaller than 0.25 per event instead of 0.4.
In order to check the stability of the results on the method
used to correct the unlike-charge distribution for K$_S^0$ and
resonance decay products, the measured resonance rates are varied
inside the experimental errors to obtain maximum and minimum values
for the factors used to correct (i.e. either subtracted or added 
according to a defect or an excess with respect to the simulated resonance
rates) $N^{MC}_{unlike}$. 
To estimate the influence of the long-range correlations, 
the fit is performed in a more restricted interval, 0.04~$\leq$~Q$_{t_{out}}$,
Q$_{t_{side}}$,Q$_l$~$\leq$~1.0 GeV.
Finally, the difference between the results of the fits to the correlation
functions C and ${\mathrm C}'$ is considered as an asymmetrical contribution
to the systematic error.
Table~\ref{systemqtl} shows the various contributions to the systematic
error.
\begin{table}[h]
\begin{center}
\footnotesize
\begin{tabular}{|l||c|c|c|c|c|}
\hline
  & $r_{t_{out}}$~[fm] & $r_{t_{side}}$~[fm] &
$r_l$~[fm] & $r_l/r_{t_{side}}$ & $\chi^2/$DoF
\cr
\hline\hline
a) Reference fit & $0.647 \pm 0.011$ & $0.809 \pm 0.009$ &
$0.989 \pm 0.011$ & $1.222\pm 0.027$ & 25398/24428
\cr
\hline
b) Modified track selection & $0.656 \pm 0.012$ & $0.815 \pm 0.009$ &
$0.995 \pm 0.012$   & $1.221 \pm 0.028$  & 25209/24428
\cr
c) Max. resonance correction & $0.639 \pm 0.010$ & $0.812 \pm 0.008$ &
$0.988 \pm 0.010$   & $1.217 \pm 0.024$ & 25566/24428
\cr
c) Min. resonance correction & $0.657 \pm 0.011$ & $0.805 \pm 0.009$ &
$0.989 \pm 0.011$   & $1.228 \pm 0.027$ & 25319/24428
\cr
d) Restricted fit range &
$0.627 \pm 0.012$ & $0.791 \pm 0.009$ & $0.975 \pm 0.011$ & $1.233 \pm 0.028$
& 13751/13053
\cr
e) Correlation function C & $0.525 \pm 0.006$ & $0.783 \pm 0.006$ &
$1.015 \pm 0.009$ & $1.296 \pm 0.021$  & 32758/24428
\cr
\hline
\end{tabular}
\caption{\sl{Results of the fit of Eq.~9 to several variations of
         ${\mathrm C'}$({\rm Q}$_{t_{out}}$,{\rm Q}$_{t_{side}}$,
 {\rm Q}$_{l}$), as listed in the text and in the first column.
         The quoted errors are only statistical.}}
\label{systemqtl}
\end{center}
\end{table}
The global systematic uncertainties are computed by
adding in quadrature the differences between the reference fit a)
and the variations b)~--~e).
\newline

The conclusion from this analysis is that, as observed in the LCMS, the pion
emitting region is elongated, with r$_l$ greater than r$_{t_{side}}$.
From Fig.~6 it is evident that the ratio r$_l$/r$_{t_{side}}$
has a (small) dependence on y$_{cut}$; the largest value is obtained 
for smaller y$_{cut}$.
One also observes that r$_l$ is independent on y$_{cut}$, while r$_{t_{side}}$
increases with increasing y$_{cut}$.
As an example, we quote the following parameter values obtained 
for two-jet events, selected using y$_{cut}$~=~0.04:
\begin{flushleft}
r$_{t_{out}}$~=~(0.647~$\pm$~0.011~({\it stat})~$^{+0.024}_{-0.124}$~({\it syst}))~{\rm fm}
\end{flushleft}
\begin{flushleft}
r$_{t_{side}}$~=~(0.809~$\pm$~0.009~({\it stat})~$^{+0.019}_{-0.032}$~({\it syst}))~{\rm fm}
\end{flushleft}
\begin{flushleft}
r$_l$~=~(0.989~$\pm$~0.011~({\it stat})~$^{+0.030}_{-0.015}$~({\it syst}))~{\rm fm}
\end{flushleft}
\begin{equation}
\end{equation}
\begin{flushleft}
r$_l$/r$_{t_{side}}$~=~1.222~$\pm$~0.027~({\it stat})~$^{+0.075}_{-0.012}$~({\it syst}).
\end{flushleft}
\vspace{5mm}

The results of this analysis are in qualitative agreement with
recent results from the L3 collaboration~\cite{L3}.
In the L3 analysis, which uses an event-mixing technique to compute
the reference sample, the ratio of transverse to longitudinal radius
is 0.81~$\pm$~0.02~$^{+0.03}_{-0.19}~$, corresponding to 
r$_l$/r$_{t_{side}}$~=~1.23~$\pm$~0.03~$^{+0.29}_{-0.05}~$.
\newline

The results can also be compared with the predictions of a recent model of 
BECs based on string fragmentation~\cite{lund_bec}.
In this model, the different mechanisms that generate the longitudinal 
(i.e. along the string) and transverse momenta of the particle, lead to a 
longitudinal correlation length,
representing the space-time difference along the string between the production
points, which is larger than the transverse correlation length.
\newline

As a check, a two-dimensional analysis where the transverse 
component Q$_t$ of the pair momentum difference is not split into ``out" 
and ``side" components, is done.
The two-dimensional Goldhaber function fitted to the correlation function 
${\mathrm C'}$(Q$_t$,Q$_l$) gives, in the case of two-jet events
selected with y$_{cut}$~=~0.04, 
r$_l$/r$_t$~=~1.360~$\pm$~0.026~({\it stat}), in agreement with a 
longitudinal source size larger than the transverse size. 
While r$_{t_{out}}$ and r$_{t_{side}}$ show a slight dependence on the jet
resolution parameter y$_{cut}$ (in opposite directions, see Fig.~6), 
the parameter r$_t$ obtained from the two-dimensional analysis is 
independent on y$_{cut}$. 
\newline\indent
We conclude that we always have r$_l$/r$_{t_{side}}$ greater than one.
%
%
\section{One-dimensional analysis of the inclusive sample}
\begin{figure}[t]
\vspace{-12mm}
\centerline{\epsfxsize=10cm
            \epsffile{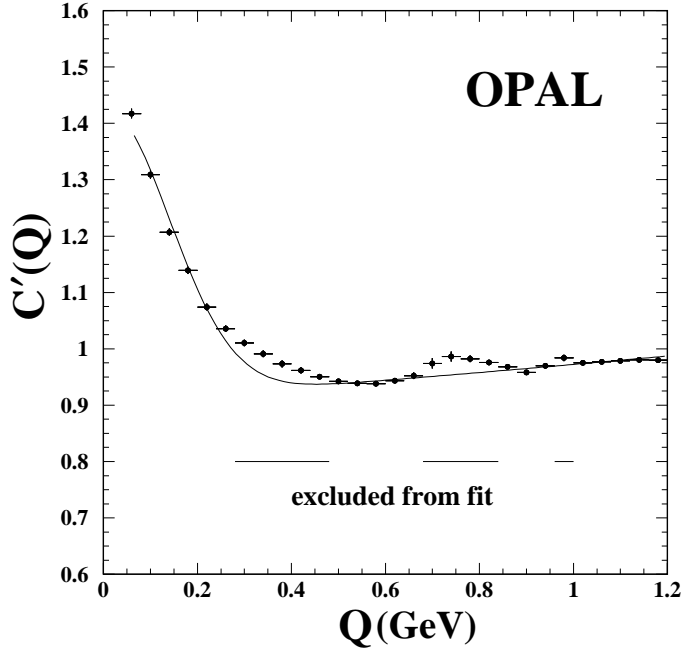}}
\caption{\sl{The one-dimensional correlation function ${\mathrm C'}$(Q) with
         the result of the one-dimensional Goldhaber fit, Eq.~13, 
         superimposed.}}
\label{q}
\end{figure}

%
%
The one-dimensional correlation functions
${\mathrm C}$(Q) and ${\mathrm C'}$(Q), where Q is the modulus of
the four-momentum difference of the pion pair, are studied for the
inclusive sample.
The results can be compared with those published in~\cite{ee_bec};
the present analysis uses more data (4.3 million Z$^0$ hadronic decays
instead of 3.6 millions) and a different version of the Jetset Monte Carlo
(7.4 instead of 7.3).
The correlation function ${\mathrm C'}$(Q), after the corrections for 
Coulomb and resonance decay products, is shown in Fig.~\ref{q}.
The one-dimensional Goldhaber function
\begin{equation}
{\mathrm C}'({\mathrm Q})=
N(1+ \lambda e^{-{\mathrm Q}^{2}r^{2}})
(1+\delta{\mathrm Q}+\epsilon{\mathrm Q}^2)
\end{equation}
is fitted to the measured correlation function over the range 
0.04~$\leq$~Q~$\leq$~1.2 GeV.
There are apparent distortions in the unlike-charge pairs distribution 
even after the correction for the contribution of resonance decays products,  
as discussed in the three-dimensional analysis.
Therefore the Q regions affected most by $\omega$, $\eta$, K$^0_S$, $\rho^0$
and $f_0(980)$ decay products, corresponding respectively to 
0.28~$\leq$~Q~$\leq$~0.48 GeV, 0.68~$\leq$~Q~$\leq$~0.84 GeV 
and 0.96~$\leq$~Q~$\leq$~1.0 GeV, are excluded from the fit range.
\begin{table}[htbp]
\begin{center}
\footnotesize
\begin{tabular}{|l||c|c|c|}
\hline
 & r [fm] & $\lambda$ & $\chi^2$/DoF
\cr
\hline\hline
a) Reference fit & 1.002 $\pm$ 0.016 & 0.574 $\pm$ 0.019 & 69/15=4.6
\cr
\hline
b) Modified track selection & 0.984 $\pm$ 0.017 & 0.590 $\pm$ 0.021 & 55/15=3.7
\cr
c) Max. resonance correction & 0.991 $\pm$ 0.016 & 0.605 $\pm$ 0.019 
& 66/15=4.4
\cr
c) Min. resonance correction & 1.014 $\pm$ 0.017 & 0.546 
$\pm$ 0.019 & 75/15=5.0
\cr
d) Restricted fit range & 1.011 $\pm$ 0.023 & 0.557 $\pm$ 0.024 & 
67/10=6.7 
\cr
e) Correlation function C & 0.909 $\pm$ 0.017 & 0.580 $\pm$ 0.022 & 150/15=10.0
\cr
\hline
\end{tabular}
\caption{\sl{Results of the fits of Eq.~13 to several variations of 
          ${\mathrm C'}$(Q). 
          The quoted errors are only statistical.}}
\label{systq}
\end{center}
\end{table}

The systematic error on the measured values of the parameters {\it r}
and $\lambda$ is evaluated in the same way as done for the
three-dimensional analysis.
The results of the various fits are summarized in Table~\ref{systq}.
\newline

The fit gives the following values for the parameters:
\begin{equation}
{\it r}~=~(1.002~\pm~0.016~({\it stat})~^{+0.023}_{-0.096}~({\it syst}))
~{\mathrm fm}~~~,~~~\lambda~=~0.574~\pm~0.019~({\it stat})~^{+0.039}_
{-0.036}~({\it syst}).
\end{equation}

Since the percentage number of charged tracks which are pions is about 87\%,
one can estimate that the value of the $\lambda$ parameter would be
a factor of 1.32 larger in the case of a 100\% pure pion sample.
\newline

The values of Eq. 14 are in agreement and replace the values previously
published by the OPAL Collaboration, see ref.~\cite{ee_bec}.
%
%
\section{Conclusions}
%
%
Using 4.3 million hadronic events from Z$^0$ decays,
the Bose-Einstein correlation function for two identical charged bosons,
mainly $\pi^{\pm}$$\pi^{\pm}$, is studied in the three components of the 
momentum difference, longitudinal and transverse (``out" and
``side" components) with respect to the 
thrust direction, in the Longitudinally CoMoving System.
The geometrical structure of the source is obtained
from an extended Goldhaber fit of Eq.~9 to the Coulomb corrected 
BEC functions
C(Q$_{t_{out}}$,Q$_{t_{side}}$,Q$_l$) and 
${\mathrm C'}$(Q$_{t_{out}}$,Q$_{t_{side}}$,Q$_l$).
In all cases the longitudinal radius is significantly larger than the
transverse radius.
\newline

The longitudinal and transverse radii of the emitting region are
studied as a function of the two-jet resolution parameter y$_{cut}$.
The analyses indicate that, as y$_{cut}$ increases, r$_{t_{side}}$ increases
slowly, r$_{l}$ remains constant and that the ratio r$_{l}$/r$_{t_{side}}$
decreases.
In the framework of the string model of ref.~\cite{lund_bec}, the observed
different values of the transverse and longitudinal correlation lengths
are explained in terms of two different generation mechanisms of the
longitudinal and transverse momentum components with respect to the
string direction.
\newline

The fit of Eq.~9 to ${\mathrm C'}$(Q$_{t_{out}}$,Q$_{t_{side}}$,Q$_l$), 
in the case of two-jet events selected with y$_{cut}$~=~0.04, 
yields the parameter values given in (12), in particular
r$_l$/r$_{t_{side}}$~=~1.222~$\pm$~0.027~({\it stat})~$^{+0.075}_{-0.012}$
~({\it syst}).
The corresponding value for the inclusive sample is 
r$_l$/r$_{t_{side}}$~=~1.194, with similar uncertainties.
\newline

The inclusive sample of events is also analysed in terms of the 
one-dimensional correlation function ${\mathrm C'}$(Q). The fit gives
the parameter values
{\it r}~=~(1.002~$\pm$~0.016~({\it stat})~
$^{+0.023}_{-0.096}$~({\it syst}))~fm and 
$\lambda$~=~0.574~$\pm$~0.019~({\it stat})~$^{+0.039}_{-0.036}$~({\it syst}).
\newline

In conclusion, the present analysis shows that the emitting source of 
two identical pions, measured in e$^+$e$^-$ interactions at energies
close to the Z$^0$ peak, is elongated.
In particular, as computed in the LCMS, the emitting source for two identical 
pions has global dimensions of about 1~fm, but with the longitudinal 
dimension about 20\% larger than the transverse dimension.
\vspace{14mm}
\newline
%
%
{\Large Acknowledgements}
%
\newline

\noindent
We particularly wish to thank the SL Division for the efficient operation
of the LEP accelerator at all energies
 and for their continuing close cooperation with
our experimental group. We thank our colleagues from CEA, DAPNIA/SPP,
CE-Saclay for their efforts over the years on the time-of-flight and trigger
systems which we continue to use. In addition to the support staff at our own
institutions we are pleased to acknowledge the  \\
Department of Energy, USA, \\
National Science Foundation, USA, \\
Particle Physics and Astronomy Research Council, UK, \\
Natural Sciences and Engineering Research Council, Canada, \\
Israel Science Foundation, administered by the Israel
Academy of Science and Humanities, \\
Minerva Gesellschaft, \\
Benoziyo Center for High Energy Physics,\\
Japanese Ministry of Education, Science and Culture (the
Monbusho) and a grant under the Monbusho International
Science Research Program,\\
Japanese Society for the Promotion of Science (JSPS),\\
German Israeli Bi-national Science Foundation (GIF), \\
Bundesministerium f\"ur Bildung, Wissenschaft,
Forschung und Technologie, Germany, \\
National Research Council of Canada, \\
Research Corporation, USA,\\
Hungarian Foundation for Scientific Research, OTKA T-029328,
T023793 and OTKA F-023259.\\
\newpage


\begin{thebibliography}{99}

\bibitem{comp_bec}
See for example
E.~A.~De~Wolf; {\it Bose-Einstein Correlations} in Proc. XXVII Int.
Conf. on High Energy Physics, Glasgow 20-27 July 1994 (Eds. P.~J.~Bussey
and I.~J.~Knowles), Inst. of Phys. Publ., 1995, p.~1281;\\
G.~Alexander and I.~Cohen; {\it Measure of $\pi$'s and $\Lambda$'s
emitter radius via Bose-Einstein and Fermi-Dirac statistics} in
Proc. Int. Conf. on Hadron Structure, Stara Lesna, Slovakia,
7-13 September 1998 (Eds. D. Bruncko and P. Strizenec) p.~328.

\bibitem{ee_bec}
OPAL Coll., G.~Alexander \etal, \ZPhys\ C72 (1996) 389.

\bibitem{ee_adl_bec}
ALEPH Coll., D.~Decamp \etal, \ZPhys\ C54 (1992) 75;\\
DELPHI Coll., P.~Abreu \etal, \PhysLett\ B286 (1992) 201.

\bibitem{ep_bec}
H1 Coll., C.~Adloff \etal, \ZPhys\ C75 (1997) 437.

\bibitem{pp_bec}
ABCDHW Coll., A.~Breakstone \etal, \PhysLett\ B162 (1985) 400;\\
UA1 Coll., C.~Albajar  \etal, \PhysLett\ B226 (1989) 410;\\
E735 Coll., T.~Alexopoulos \etal,  Phys.~Rev. D48 (1993) 1931.

\bibitem{pikp_bec}
EHS/NA22 Coll., N.M.~Agababian \etal, \ZPhys\ C71 (1996) 405;\\
EHS/NA22 Coll., N.M.~Agababian \etal, \PhysLett\ B422 (1998) 359.

\bibitem{hi_bec}
NA35 Coll., T.~Alber \etal,  \ZPhys\ C66 (1995) 77;\\
NA44 Coll., I.G.~Bearden \etal, Phys.~Rev. C58 (1998) 1656.

\bibitem{kk_bec}
ALEPH Coll., D.~Buskulic \etal, \ZPhys\ C64 (1994) 361;\\
DELPHI Coll., P.~Abreu \etal, \PhysLett\ B323 (1994) 242;\\
OPAL Coll., R.~Akers \etal, \ZPhys\ C67 (1995) 389.

\bibitem{kpkm_bec}
DELPHI Coll., P.~Abreu \etal, \PhysLett\ B379 (1996) 330;\\
OPAL Coll., G.~Abbiendi \etal, CERN EP-99/163, submitted to Eur.~Phys.~Jou.~C.

\bibitem{ww_bec}
DELPHI Coll., P. Abreu \etal, \PhysLett\ B401 (1997) 181;\\
OPAL Coll., G.~Abbiendi \etal, Eur.~Phys.~Jou.~C8 (1999) 559.

\bibitem{trepi}
DELPHI Coll., P.~Abreu \etal, \PhysLett\ B355 (1995) 415;\\
OPAL Coll., K.~Ackerstaff \etal, Eur.~Phys.~Jou.~C5 (1998) 239.

\bibitem{theor}
U.~Heinz \etal, \PhysLett\ B382 (1996) 181;\\
U.~Heinz; {\it Hanbury-Brown/Twiss interferometry for relativistic
heavy ion collisions: theoretical aspects}, nucl-th/9609029;\\
K.~Geiger, J.~Ellis, U.~Heinz and U.A.~Wiedemann;
{\it Bose-Einstein Correlations in a space-time approach to
e$^+$e$^-$ annihilation into hadrons}, hep-ph/9811270.

\bibitem{lund_bec}
B.~Andersson and M.~Ringner, \NPhys\ B513 (1998) 627;
\PhysLett\ B421 (1998) 283.

\bibitem{tasso}
TASSO Coll., M.~Althoff \etal, \ZPhys\ C30 (1986) 355.

\bibitem{L3}
L3 Coll., M.~Acciarri \etal, \PhysLett\ B458/4 (1999) 517.

\bibitem{opaldete}
OPAL Collaboration, K.~Ahmet et~al., Nucl.~Instr.~and Methods~A305 (1991) 275;\\
P.P.~Allport et~al., Nucl.~Instr.~and Methods~A324 (1993) 34;
Nucl.~Instr.~and Methods~A346 (1994) 476.

\bibitem{si}
B.E.~Anderson et~al., IEEE Transactions on Nuclear Science 41 (1994) 845.

\bibitem{durham}
S.~Catani \etal, \PhysLett\ B269 (1991) 432.

\bibitem{rlcms}
T.~Cs\"{o}rg\"{o} and S.~Pratt, in Proc. of the Workshop on
Relativistic Heavy Ion Physics at Present and Future Accelerators,
Budapest, KFKI-1991-28/A 14-21, p.~75.

\bibitem{scotto}
S.~Chapman, P.~Scotto and U.~Heinz, Phys.~Rev.~Lett.~74 (1995) 4400.

\bibitem{gamow}
M.~Gyulassy \etal, Phys.~Rev.~C20 (1979) 2267.

\bibitem{Jetset}
T.~Sj\"{o}strand, Comp. Phys. Comm. 39 (1986) 347;\\
T.~Sj\"{o}strand and M Bentgtsson, Comp. Phys. Comm. 43 (1987) 367;
Comp. Phys. Comm. 82 (1994) 74.

\bibitem{lafferty}
I.G.~Knowles and G.D.~Lafferty, J.~Phys.~G23 (1997) 731.

\bibitem{gold}
G.~Goldhaber, S.~Goldhaber, W.~Lee and A.~Pais, Phys.~Rev.~Lett.~3 (1959) 181.

\end{thebibliography}
\end{document}